\documentclass[12pt]{iopart}
\usepackage{graphicx}
\usepackage{iopams}
\usepackage{color}

\begin{document}
\title[Stochastic modeling of gene expression: application of ensembles of trajectories]{Stochastic modeling of gene expression: application of ensembles of trajectories}
\author{Pegah Torkaman and Farhad H. Jafarpour\footnote{Corresponding author: farhad@ipm.ir}}
\address{Physics Department, Bu-Ali Sina University, 65174-4161 Hamedan, Iran}

\begin{abstract}
It is well established that gene expression can be modeled as a Markovian stochastic process and hence proper observables 
might be subjected to large fluctuations and rare events. Since dynamics is often more than statics, one can work with ensembles 
of trajectories for long but fixed times, instead of states or configurations, to study dynamics of these Markovian stochastic processes 
and glean more information. In this paper we aim to show that the concept of ensemble of trajectories can be applied to a variety of 
stochastic models of gene expression ranging from a simple birth-death process to a more sophisticate model containing burst and switch. 
By considering the protein numbers as a relevant dynamical observable, apart from asymptotic behavior of remote tails of probability 
distribution, generating function for the cumulants of this observable can also be obtained. We discuss the unconditional stochastic Markov processes
which generate the statistics of rare events in these models. 
\end{abstract}

\maketitle
%%%%%%%%%%%%%%%%%%%%%%%%%%%%%%%%%%%%%%%%%%%%%%%%%%%%%%%%%%%%%%%%%%%%%%%%%
\section{Introduction}
%%%%%%%%%%%%%%%%%%%%%%%%%%%%%%%%%%%%%%%%%%%%%%%%%%%%%%%%%%%%%%%%%%%%%%%%%
Traditionally, stochastic simulation has been a powerful tool for studying the dynamics of gene regulatory networks~\cite{simul}. However, this approach 
is not always efficient specially when rare events occur~\cite{simulrare}. 

The ergodic theorem justifies the replacement of a long-time average of any physical quantity, or equivalently a laboratory measurement, with an 
ensemble average of that quantity overt configurations. However, when it the comes to the dynamics of the system when a rare value of a 
time-integrated quantity (time-dependent observables) is observed, the concept of the ensemble of trajectories can serve well. In this case one is interested in studying rare 
trajectories during which the system behaves in an unusual fashion over an extended period of time and time-averaged quantities remain far 
from their typical values. While it is hard to observe such rare events experimentally, because of having an exponentially small probability of 
occurrence, they can be studied using the theory of large deviations. 

The theory of large deviations deals with the probabilities of rare events~\cite{T09}. This theory describes how the probability of occurrence 
of a  time-integrated dynamical observable which is largely deviated from its typical value decays to zero as the length of trajectory goes to infinity. 
On the other hand, the dynamics of stochastic systems is usually richer than what can be gleaned from their stationary properties. 
In order to fully understand dynamical behavior of the system, one necessarily has to focus on trajectories, and not just on configurations. 
A possible efficient approach for achieving that is to study the statistics of dynamical observables whose fluctuation behavior characterizes 
the dynamics of the system. 

The time-integrated/averaged quantities and their fluctuations are of biological significance~\cite{BP1977}.
Given the fact that gene expression is intrinsically a stochastic process, fluctuations or rare events of proper time-integrated quantities 
might be observed in this process. Several prominent examples of such rare events have been observed in biological systems which include phenotypic 
transitions~\cite{phtrans}. Cumulative sum of proteins have also been considered in modeling approaches dealing with cancer 
therapy~\cite{BBS2018}; therefore, fluctuations of such qualities could be interesting from biological point of view.

Recently, a vast amount of research is devoted to show that the dynamics of 
stochastic systems can be studied by analyzing the statistical properties of dynamical trajectories 
(time-realizations)~\cite{S09,PSS10,BS13,LW12,PH14,ZRP,Gising,Gglass,G1order}. 
It is of great importance to characterize fluctuations in the system conditioned on the occurrence of a rare event.
Ensembles of trajectories are associated with large deviations of time-integrated quantities. 
While some trajectories are responsible for creating typical values of the observable, some other trajectories create a specific fluctuation or an 
atypical value of the observable. By looking at a certain set of trajectories which is responsible for a specific fluctuation, we are imposing  
constraints on the time-integrated quantities~\cite{TCh15}. It has been shown that one can determine the changes in dynamical model parameters so 
that it reproduces the effects of rare fluctuations~\cite{JS10}. Far from the boundaries of the observation time, there is a temporal regime during which
the trajectories of the biased ensemble are close to the steady-state trajectories of an auxiliary model (sometimes called driven or effective model) 
which is a conditioning-free stochastic Markov process. In other words, the statistics of the conditioning-free Markov stochastic process reproduces the 
fluctuations of the original Markov process conditioned on the occurrence of a rare event~\cite{JS15}.

In comparison with the static ensemble of configurations approach, the above mentioned approach provides us, in principle, with a variety of information. 
Assuming that the probability distribution function of the dynamical observable has a large deviation form, one can calculate the large deviation function. 
The scaled cumulant generating function of the time-integrated observable can also be calculated. It is also possible to determine the probability vectors of both the 
final and initial configuration, knowing that the value of the observable through the evolution of the system is restricted 
to a certain value. Last but not least, the effective process can also be calculated~\cite{T09,G1order,JS10}.

In this paper we aim to show that using the ensemble of trajectories approach one can obtain a better understanding of the dynamical behavior of the 
stochastic models of gene expression. After a brief review of Mathematical preliminaries we start with the simplest model of gene expression consisting of
Poissonian creation and degradation of proteins. We will then add burst and switch of promoter to the models and show that these models are still solvable.  
In comparison with a recent work on modeling of stochastic gene expression conditioned on large deviations~\cite{HK17}, we will show that it is possible to add
degradation of proteins/mRNA's to their model and, apart from what have been calculated, one can answer quite interesting questions such as
which dynamical trajectories have the most contribution to occurrence of a rare event.   

%%%%%%%%%%%%%%%%%%%%%%%%%%%%%%%%%%%%%%%%%%%%%%%%%%%%%%%%%%%%%%%%%%%%%%%%%
\section{Mathematical preliminaries}
%%%%%%%%%%%%%%%%%%%%%%%%%%%%%%%%%%%%%%%%%%%%%%%%%%%%%%%%%%%%%%%%%%%%%%%%%
Let us consider a stochastic process system with discrete configuration space $\{ c \}$. The time evolution equation for probability of being in 
configuration $c$ at time $t$, denoted by $P(c,t)$, is given by 
\begin{equation}
\label{me}
\frac{d}{dt} P(c,t)=\sum_{c'\ne c} w_{c' \to c} P(c',t)-\sum_{c' \ne c} w_{c\to c'} P(c,t) 
\end{equation}
in which $w_{c\to c'}$ is a time-independent rate of jumping from configuration $c$ to $c'$. By introducing an orthonormal 
complete basis $\{ \vert c \rangle \}$ with $\langle c \vert c' \rangle =\delta_{c,c'}$ and defining
$$
 \langle c \vert P(t) \rangle \equiv P(c,t)
$$
and the matrix elements of $H$ as
$$
\langle c \vert H \vert c' \rangle \equiv w_{c' \to c} \;\; \mbox{for} \;\; c' \neq c \;\; \mbox{and}\;\;
\langle c \vert H \vert c \rangle\equiv -\sum_{c' \neq c}w_{c \to c'}\;\; \mbox{for} \;\; c' = c 
$$
we can rewrite the Master equation~(\ref{me}) as follows~\cite{Schbook} 
\begin{equation}
\frac{d}{dt} \vert P(t) \rangle= H \vert P(t) \rangle 
\end{equation}
in which $\vert P(t) \rangle $ is the probability vector at time $t$ and $H$ is the stochastic generator of the process with the following property 
$$
\langle 1 \vert H=0  
$$
where we have defined a summation vector $\langle 1 \vert$ as $\langle 1 \vert \equiv \sum_{c} \langle c \vert$. In the long time limit
the steady-state probability vector satisfies
$$
H\vert P^{\ast}\rangle =0 \; .
$$

Instead of looking at the time-evolution of the probability vector and calculating the average values of the observables over a static ensemble of configurations, we 
can look at realizations or trajectories of the process in the space of configurations of the system. By defining dynamical observables over these 
trajectories, we can then study the dynamical properties of the process such as dynamical phase transitions~\cite{TCh15,Ltraj}. 
There are usually two types of dynamical observables or equivalently time-integrated observables. The first type is a purely spatial observable 
and depends on both the configurations that the system meets along a trajectory and the time it spends in each configuration. This type of a 
time-integrated observable which is a functional of the trajectory can be written as 
$$
\sum_{i} (t_{i+1}-t_{i}) \phi_{c(t_i)}
$$
in which $\phi_{c(t_i)}$ is the increment of the observable when it meets the configuration $c$ at time $t_i$. For example, $\phi _c (t_i)$ can be density 
of particles or energy of the system associated with configuration $c$ at time $t_i$. The second type of a time-integrated observable 
depends on the transitions among consecutive configurations along a trajectory and can be written as
$$
\sum_{i} \theta_{c(t_i)\to c(t_{i+1})}
$$
in which the increment of this current is denoted by $\theta_{c(t_i)\to c(t_{i+1})}$. This kind of dynamical observable is sometimes called current-like 
observable, such as particle current, entropy production rate and activity. For example, taking $\theta_{c(t_i) \to c(t_{i+1})}=1$ for all transitions led to 
counting the number of configuration changes during the observation time which is called the activity of a system~\cite{T09,JS10}. 
We can build an ensemble of trajectories in different ways as we do in traditional classical statistical mechanics for the ensemble of configurations. 
This will be discussed in the following. 

We start with the micro-canonical ensemble of configurations defined for an isolated system with fixed energy $E$.
The constraint of fixed energy means that every member of the micro-canonical ensemble has the same energy. It is exactly 
because of this constraint that calculating the conditional probability of being in $c$ given that $E$ is fixed $P(c\vert E)$ might be difficult.
Similarly, but not on the same footing, we can consider an ensemble of trajectories, defined between an initial time $0$ and a final 
time $t$ (called the observation time which is assumed to be very long), with a constraint on the value of a given dynamical observable 
over the trajectories. This ensemble contains those trajectories, among all possible trajectories, for which the value of that observable is fixed along the trajectory~\cite{G1order,TCh15,HSch07}. 

We can also consider an ensemble of trajectories assuming that the average value of a given dynamical observable is constant during the observation 
time~\cite{Ltraj,HSch07}. This is similar to the canonical ensemble of configurations in the traditional statistical mechanics where the probability of being in a configuration 
is calculated at fixed temperature (i.e. the average value of the energy is fixed). This can be achieved by multiplying the probabilities in the micro-canonical 
ensemble by $e^{-\beta H(c)}$ in which $H(c)$ is the Hamiltonian of the system while being in the configuration $c$. As the temperature is varied, the average 
energy is changed. Calculating the average value of any observables in this canonical ensemble of energy gives the typical value of those 
observables under fixed temperature. For the ensemble of trajectories this can be formulated as follows: We multiply the probability of taking the trajectory 
$\cal C$, denoted by $P[{\cal C}]$, by $e^{-s{\cal A}_t[{\cal C}]}$ in which ${\cal A}_t$ is a time-extensive dynamical observable explained before. 
$s$ is a Mathematical biasing field conjugated to $\mathcal{A}_t$, analogous to parameter $\beta$ in traditional statistical physics. 
The ensemble average of a dynamical observable, say ${\cal O}_t[{\cal C}]$, which is a functional of $\cal C$ is given by 
\begin{equation}
\label{avg}
\langle {\cal O} \rangle_s =\frac{1}{Z} \sum_{\cal C}  {\cal O}_t[{\cal C}] P[{\cal C}] e^{-s{\cal A}_t[{\cal C}]} \; .
\end{equation}
The dynamical partition function $Z$ appeared in~(\ref{avg}) is a normalization factor whose logarithm is called the dynamical free energy. 
Note that, as in the canonical ensemble, the field $s$ conjugated to ${\cal A}_t$ can fix the average value of this observable during the observation time. 
This means that the averages are now calculated in an ensemble of trajectories under fixed $s$. The above statistics is constructed over a biased ensemble 
of trajectories which is sometimes called the $s$-ensemble.
Averages in the $s$-ensemble with $s = 0$ correspond to the steady-state averages of ${\cal O}_t$ which is the only physically accessible ensemble.
These averages are clearly time-translational invariant; however, this invariance is broken for the averages over the $s$-ensemble with $s\neq 0$. 
If the observation time is very long and we are far from the boundaries $t'=0$ and $t'=t$, there is a temporal regime  
during which the time translational invariance (TTI) is recovered. In other words, the $s$-ensemble averages during 
the TTI regime do not depend on the instance of time at which the averaging in performed. However, before the system relaxes into the TTI regime
and after it goes out of it, these averages depend on the time of averaging~\cite{G1order,JS10}. 

Let us look at the large deviations of the dynamical observable ${\cal A}_t$ in the long time limit. The generating function of the moments of the 
time-integrated quantity $a_t=\frac{1}{t} {\cal A}_t$ can be written as~\cite{G1order}
\begin{equation}
\label{scgf1}
Z=\langle e^{-t s a_t} \rangle_s=\langle 1 \vert e^{t H(s)} \vert P(0)\rangle
\end{equation}
in which the average $\langle e^{-t s a_t} \rangle_s$ is taken over all possible trajectories during the time interval $[0,t]$ and $H(s)$ is a modified generator which is equal to the
stochastic generator of the process $H$ at $s=0$. The matrix elements of $H(s)$ are
$$
\langle c \vert H(s) \vert c' \rangle=w_{c' \to c} e^{-s \theta_{c'\to c}} \;\; \mbox{for} \;\; c' \neq c \;\;\mbox{and}\;\;
\langle c \vert H(s) \vert c \rangle=-\sum_{c' \neq c}w_{c \to c'}-s\phi_{c} \;\; \mbox{for} \;\; c' = c
$$
in which $\phi_c$ and $\theta_{c' \to c}$ are the increments of the time-dependent observables of the first and the second type respectively from which we construct the 
$s$-ensemble~\cite{T09, TCh15}. In the long time limit one can write 
\begin{equation}
\label{scgf2}
\lim_{t \to \infty}\frac{1}{t}\ln \langle e^{-t s a_t} \rangle_s=\Lambda^{\ast}(s)
\end{equation}
where $\Lambda^{\ast} (s)$, which is called the Scaled Cumulant Generating Function (SCGF) of the observable, is the largest eigenvalue of the modified generator $H(s)$. 
Note that the derivatives of $\Lambda^{\ast} (s)$ evaluated at $s = 0$ give the cumulants of ${\cal A}_t$ scaled by time i.e.
\begin{equation}
\label{SCF}
\lim_{t \to \infty} \frac{1}{t} \langle {\cal A}_t \rangle_c = (-1)^n \frac{d^n}{ds^n} \Lambda^{\ast}(s) \Big \vert_{s=0}
\end{equation}
from which one can characterize the fluctuations of the observable. According to the G\"artner–Ellis Theorem if $\Lambda^{\ast} (s)$ exists and is 
differentiable for all $s \in \mathbb{R}$, then $a_t$ satisfies a large deviation principle namely 
$$
P(a_t \in da) \asymp e^{-t I(a)} da
$$
with the rate function $I(a)$ given by the Legendre–Fenchel transform of $\Lambda^{\ast} (s)$
\begin{equation}
\label{ld}
I(a)=-\inf_{s \in \mathbb{R}} \{ sa+\Lambda^{\ast} (s)  \}
\end{equation}
in which the symbol ‘‘inf’’ above stands for ‘‘infimum of’’, which can be taken to mean the same as ‘‘minimum of’’. Expanding $I(a)$ beyond quadratic order 
gives information about non- Gaussian fluctuations of ${\cal A}_t/t$, which are referred to as large deviations. Moreover, from~(\ref{ld}) we find
\begin{equation}
\label{mv}
a=\langle a_t \rangle_s=-\frac{d}{ds} \Lambda^{\ast}(s)
\end{equation}
a relation which connects the average value of the observable $a_t$ to the conjugate field $s$~\cite{T09}. The right and the left 
eigenvectors of $H(s)$ corresponding to the largest eigenvalue of the modified generator $\Lambda^{\ast}(s)$ are quite meaningful in the context of large 
deviations in terms of trajectories. Considering the following eigenvalue relations 
\begin{eqnarray} 
H(s) \vert \Lambda^{\ast} \rangle= \Lambda^{\ast}(s)  \vert \Lambda^{\ast} \rangle \;,\\
\langle \tilde\Lambda^{\ast} \vert H(s) = \Lambda^{\ast}(s)  \langle \tilde\Lambda^{\ast} \vert
\end{eqnarray}
it has been shown that 
\begin{eqnarray}
\vert \Lambda^{\ast} \rangle \propto \sum_{c} P_f(c ,s) \vert c \rangle \;, \\
\vert \tilde \Lambda^{\ast} \rangle = {\langle \tilde\Lambda^{\ast} \vert}^T \propto \sum_{c} P_i(c ,s)  \vert c \rangle
\end{eqnarray}
in which $P_f(c ,s)$ and $P_i(c ,s)$ are in fact the probability of the final and initial configuration respectively, knowing that the value of $a$ is fixed and related 
to the observable $s$ through~(\ref{mv})~\cite{S09,G1order,JS10}. It is worth mentioning that these normalized probabilities are defined as follows
\begin{eqnarray}
\label{final}
P_f(c ,s) = \frac{\langle c \vert \Lambda^{\ast} \rangle}{\langle 1 \vert \Lambda^{\ast} \rangle} \; ,\\
\label{initial}
P_i(c ,s) =\frac{\langle \tilde \Lambda^{\ast} \vert c \rangle \langle c \vert P(0) \rangle}{\langle \tilde \Lambda^{\ast} \vert P(0) \rangle}\; .
\end{eqnarray}
We could also look at the probability of observing a given configuration $c_{t'}$ at a time $t'$ during the evolution of the system far from the initial 
and final configuration, i.e. $0 \ll t' \ll t$, conditioned on fixed $s$. Starting from $t'=0$ the system relaxes, on a time scale $\tau$, into the TTI regime 
during which the probability of being in the configuration $c$ at fixed $s$ is given by
\begin{equation}
\label{tti}
P_{\mbox{\tiny TTI}}(c,s)=\frac{\langle \tilde \Lambda^{\ast} \vert c \rangle \langle c \vert \Lambda^{\ast} \rangle}{\langle \tilde \Lambda^{\ast} \vert  \Lambda^{\ast} \rangle}.
\end{equation}
During the TTI regime defined as $\tau \ll t', \tau \ll t-t'$ in which $\tau$ is a relaxation time into this temporal regime, the steady-state trajectories of the effective 
process, discussed below, are those of the biased ensemble of trajectories~\cite{TCh15,JS10,JS15}.

As we mentioned before, in the long time limit each specific fluctuation in the system can be described by a stochastic 
Markov process called the effective process. This process is equivalent to the conditioning of the original process on
seeing a certain fluctuation. The stochastic generator of this effective stochastic process is given by
\begin{equation}
\label{effective relation}
H_{\mbox{\tiny eff}}(s) =U(s) H(s) U^{-1}(s)-\Lambda^{\ast}(s) 
\end{equation}
which is a generalization of Doob's h-transform~\cite{TCh15,JS10}. $U(s)$ in~(\ref{effective relation}) is a diagonal matrix with the matrix element 
$ \langle c \vert U(s) \vert c \rangle = \langle \tilde{\Lambda}^{\ast} \vert c \rangle $. 
The off-diagonal matrix elements of~(\ref{effective relation}) are given by
\begin{equation}
\label{offelement}
\langle c' \vert  H_{\mbox{\tiny eff}}(s) \vert c \rangle =\langle c' \vert H(s) \vert c \rangle \frac{
 \langle \tilde{\Lambda}^{\ast} \vert c' \rangle}
{ \langle \tilde{\Lambda}^{\ast} \vert c \rangle} \; .
\end{equation}
and the diagonal elements can be obtained using the fact that~(\ref{effective relation}) is a stochastic matrix~\cite{JS10}. 

Finally, there is an important family of stochastic processes for which the effective process is identical to the original 
process except that the effective rates are just rescaled values of those in the original process~\cite{TJ15}. This property has an important 
consequence. If we can calculate the the effective rates (by calculating the left eigenvector $\langle \tilde \Lambda^\ast \vert$) and 
the steady-state probability vector $\vert P^\ast \rangle$, we can substitute the original rates in the 
steady-sate probability vector with the effective rates and using~(\ref{tti}) obtain the right eigenvector $\vert \Lambda^\ast \rangle$ of the modified generator
associated with $\Lambda^\ast(s)$. We will use this property later  in the paper.

Considering this short introduction to the large deviations, we apply the above results to different well-known models of gene expression. 
We will start with the simplest model and then cover the more detailed ones. We will show how the powerful toolbox of large deviations 
can serve us find considerable amount of information about the dynamics of these models.      

%%%%%%%%%%%%%%%%%%%%%%%%%%%%%%%%%%%%%%%%%%%%%%%%%%%%%%%%%%%%%%%%%%%%%%%%%
\section{The models}
%%%%%%%%%%%%%%%%%%%%%%%%%%%%%%%%%%%%%%%%%%%%%%%%%%%%%%%%%%%%%%%%%%%%%%%%%
As we mentioned earlier, gene expression is intrinsically a stochastic process. After modeling of gene expression as a Markovian stochastic process 
one can, in principle, calculate the probability distribution of many relevant observables, such as number of proteins or mRNA's, and also their 
moments~\cite{SGE,ShR08}. However, until recently, not much is done on the dynamical properties of this stochastic process when these observables 
are considered as a dynamical observables. 

The toolbox of large deviations, as explained briefly in the previous section, allows one to study the dynamical properties of the stochastic processes 
as well as the large deviations of these observables when considered as dynamical variables defined on trajectories or time-realizations of the process. 
This approach, as we will see, provides us with a deeper understanding of the dynamical properties of gene expression as a Markovian stochastic process. 

In what follows, we will consider stochastic models of gene expression at mRNA or protein level and investigate the dynamical properties
of these models using the concept of $s$-ensemble. We will show that the large deviation form of the probability distribution of dynamical 
observables and also the cumulants of these observables can be calculated exactly.

%%%%%%%%%%%%%%%%%%%%%%%%%%%%%%%%%%%%%%%%%%%%%%%%%%%%%%%%%%%%%%%%%%%%%%%%%
\subsection{Model 1}
%%%%%%%%%%%%%%%%%%%%%%%%%%%%%%%%%%%%%%%%%%%%%%%%%%%%%%%%%%%%%%%%%%%%%%%%%

%%%%%%%%%%%%%%%%%
\begin{figure}[t]
\centering
\includegraphics[scale=1.5]{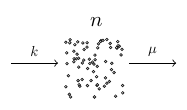}
\caption{\label{model1} A simple sketch of Model 1. }
\end{figure}
%%%%%%%%%%%%%%%%% 

As the first model we consider a birth-death model at protein level in which the proteins are created with the rate $k$ and degradation is 
occurred with the rate $n\mu$ in which $n$ is the number of the proteins~\cite{Wbook}. A simple sketch of this model is given in Fig.~(\ref{model1}). 
The Master equation for this process can be simply written as
\begin{equation}
\label{me1}
\frac{d}{dt}P(n,t)=k P(n-1,t) + \mu(n+1)P(n+1,t)-(k+n\mu)P(n,t) 
\end{equation}
in which $P(n,t)$ is the probability of existing $n$ protein at time $t$. Note that here the configuration of the system is determined by the number of proteins
and that the space of configurations is infinite dimensional. The steady-state probability distribution $P^{\ast}(n)$ (which turns to be an equilibrium 
steady-state in this case) can easily be obtained by letting $t \to \infty$ and solving~(\ref{me1}). The result is given by 
\begin{equation}
\label{ss1}
P^{\ast}(n)=e^{-\gamma} \frac{\gamma ^n}{n!} \;\;\;\mbox{with}\;\;\;  n=0,1,2,3,\dots
\end{equation}
in which $\gamma \equiv \frac{k}{\mu}$. The average number of proteins in the steady-state (which is sometimes called the typical value) can be easily calculated 
$$
\langle n \rangle=\sum_{n=0}^{\infty}n P^{\ast}(n)=\gamma \;.
$$
Since mean and noise of protein numbers are quite relevant quantities in the realm of gene expression, let us consider the number of proteins 
as a dynamical observable (the first type) defined on the trajectories or time-realizations of the process during an observation
time of length $t$. More precisely, we consider the first type time-integrated quantity $\mathcal{A}_t$ as $N_t=\int_0^t n(t) dt$ 
and define the time-averaged number of protein over observation time $t$, i.e. $n_t=N_t/t$, as the dynamical observable.
The modified generator $H(s)$ in the complete orthonormal basis $\{ \vert 0 \rangle,\vert 1 \rangle,\vert 2 \rangle,\vert 3 \rangle,\dots \}$ is given by 
\begin{equation}
\label{mg1}
H(s)=
\left( \begin{array}{ccccccc}
-k   & \mu          &     0            &          0         &\cdots\\
k    & -k-\mu-s   &  2\mu         &          0         &\cdots\\
0    &     k          &-k-2\mu-2s  &        3\mu     &\cdots\\
0    &     0          &       k          &   -k-3\mu-3s &\cdots\\
\vdots& \vdots    &    \vdots    &      \vdots      &\ddots
\end{array} \right)\, .
\end{equation}
It turns out that the full spectrum of the modified generator~(\ref{mg1}) can be calculated exactly. Considering the following eigenvalue relations
\begin{eqnarray*}
H(s) \vert \Lambda  \rangle = \Lambda (s) \vert \Lambda  \rangle \;, \\
\langle \tilde{\Lambda}  \vert H(s) = \Lambda (s) \langle \tilde{\Lambda}  \vert
\end{eqnarray*}
we have found 
\begin{eqnarray}
 \vert \Lambda_j  \rangle = \sum_{n=0}^{\infty} \phi_n^j \vert n \rangle \;, \\
\langle \tilde{\Lambda}_j \vert= \sum_{n=0}^{\infty} \tilde{\phi}_n^j \langle n \vert
\end{eqnarray}
and
\begin{equation}
\label{eigen1}
\Lambda_j (s)=-(\mu+s)j-\frac{ks}{\mu+s}   \;\;\;\mbox{with}\;\;\;  j=0,1,2,3,\dots
\end{equation}
where
\begin{eqnarray*}
 \phi_n^j=\frac{1}{n!}  C_n \Big(j ; \frac{k \mu}{(\mu+s)^2}\Big) 
 \Big(\frac{-\mu}{\mu+s}\Big)^j   \Big(\frac{k}{\mu+s}\Big)^n e^{-\frac{\mu k}{2(\mu+s)^2}}  \; , \\
\tilde{\phi}_{n}^j=\frac{1}{j!} C_n \Big(j ; \frac{k \mu}{(\mu+s)^2}\Big)  
\Big(\frac{-k}{\mu+s}\Big)^j  \Big(\frac{\mu}{\mu+s}\Big)^n e^{-\frac{\mu k}{2(\mu+s)^2}}  \;.
\end{eqnarray*}
$C_n(j;x)$ is the Poisson-Charlier polynomial defined explicitly by~\cite{charly}
\begin{equation}
C_n (j ;x)=\sum_{l=0}^{\min(n,j)} (-1)^l   
 \left(\begin{array}{c}n \\l\end{array}\right)
 \left(\begin{array}{c}j \\l\end{array}\right) l! x^{-l} 
\end{equation}
for $x>0$. These polynomials are a family of orthogonal polynomials satisfying the following relation
\begin{equation}
\sum_{j=0}^{\infty} \frac{x^j}{j!} C_n(j;x)C_m(j;x)=x^{-n}e^{x}n!\delta_{n,m} \; .
\end{equation}
Note that in~(\ref{eigen1}) $s$ is restricted to $s \in ]-\mu , +\infty]$. On the other hand, $s \in ]-\mu , 0 [$ ($s \in ]0,+\infty]$) correspond to the ensemble
average of the observable larger (smaller) than its typical value while $s=0$ gives the typical value of observable. Note that using 
$\sum_{j=0}^{\infty} \phi_m^j\tilde\phi_n^j=\delta_{m,n}$, the left and right eigenvectors of $H(s)$ satisfy
$$
\sum_{j=0}^{\infty}\vert \Lambda_j \rangle \langle \tilde \Lambda_j \vert =1 \; .
$$

The largest eigenvalue of~(\ref{mg1}) is given by $j=0$ in~(\ref{eigen1}) i.e. $\Lambda^{\ast}(s)=\frac{-ks}{\mu+s}$. As we mentioned 
earlier, derivatives of $\Lambda^{\ast}(s)$ with respect to $s$ generate the cumulants of $n_t$ in the $s$-ensemble. The first  derivative of the largest 
eigenvalue at $s=0$ gives
$$
\langle n_t \rangle_{s=0} = - \frac{d}{ds}\Lambda^{\ast}(s) \Big\vert_{s=0}=\gamma\; .
$$
According to~(\ref{SCF}), the higher-order derivatives of $\Lambda^\ast (s)$ at $s=0$ give higher-order cumulants of $N_t$ scaled with time $t$.
It is worth mentioning that while the first cumulant of the observable in the steady-state is equal to the first cumulant of the dynamical 
observable at $s=0$, the higher-order cumulants obtained from steady-state distribution of the observable are not equal to the dynamical 
ones at $s=0$. The reason is that the definition of these observables are basically different.

The right and the left eigenvectors of~(\ref{mg1}) corresponding to the largest eigenvalue $\Lambda^{\ast}(s)$ (or $j=0$ in~(\ref{eigen1})) are now given by 
\begin{equation}
 \vert \Lambda ^{\ast} \rangle = \sum_{n=0}^{\infty} \frac{1}{n!}  \Big(\frac{k}{\mu+s}\Big)^n \vert n \rangle
\end{equation}
and
\begin{equation}
\langle \tilde{\Lambda} ^{\ast} \vert= \sum_{n=0}^{\infty} \Big(\frac{\mu}{\mu+s}\Big)^n \langle n \vert 
\end{equation}
respectively. 

Let us now investigate the effective dynamics of the process. As we have already explained, an unconditional stochastic process for which
the typical value of the dynamical observable is an atypical value of the same observable in the original process, is called the effective process. 
Considering the number of proteins as a dynamical observable, the matrix elements of the stochastic generator of this effective process are found to be
\begin{eqnarray}
\langle n+1 \vert H_{\mbox{\tiny eff}}(s) \vert n \rangle &=& \frac{k \mu}{\mu +s} \; , \\
\langle n-1 \vert H_{\mbox{\tiny eff}}(s) \vert n \rangle &=& n (\mu +s) \; .
\end{eqnarray}
We remind the reader that the configuration of the system $c$ is now defined by the number of proteins in the system $n$. 
As can be seen these effective rates of birth and death are just rescaled values of those in the original process. 

It turns out that in the long time limit, the probability distribution for the number of proteins $n_t$ has a large deviation form 
$P(n_t=\bar{n})\asymp e^{-t I(\bar{n})}$ with the following rate function
\begin{equation}
I(n_t=\bar{n})=-\inf_{s} \{ \bar{n} s+\Lambda^{\ast}(s) \} = (\sqrt{k}-\sqrt{\bar{n} \mu})^2 \; .
\end{equation}

Let us have a look at another interesting quantity that is the $s$-ensemble average of protein numbers at time $t'$, a 
time between $0$ and $t$. This quantity is given by
\begin{equation}
\langle n (t') \rangle_s =
\frac{\langle 1 \vert e^{(t-t') H(s)}
\hat{n}
e^{t' H(s)} \vert \tilde{P}(s,0) \rangle}
{\langle 1 \vert \tilde{P}(s,t) \rangle }
\end{equation}
in which the modified generator $H(s)$ is given by~(\ref{mg1}) and $\tilde{P}(n,s,t') \equiv \langle n \vert \tilde{P}(s,t') \rangle$ is 
the solution of the following Master equation
\begin{eqnarray}
\frac{d}{dt'}\tilde{P}(n,s,t')&=&k \tilde{P}(n-1,s,t') + \mu(n+1)\tilde{P}(n+1,s,t') \nonumber \\
                                      &-&(k+n(\mu+s))\tilde{P}(n,s,t') \; .
\end{eqnarray}
The diagonal matrix $\hat n$ has the diagonal elements $\langle n \vert \hat n \vert n \rangle = n$. Straightforward calculations result in
\begin{equation}
\label{P(n,s,t)}
\tilde{P}(n,s,t')=e^{\psi (t')-\xi (t')} \frac{\xi (t')^n}{n!}
\end{equation}
in which
\begin{eqnarray*}
\xi (t')=\frac{k \left(s e^{-t' (\mu +s)}+\mu\right)}{\mu(\mu +s)} ,\;\;
\psi (t')=\frac{k s \left(s e^{-t' (\mu +s)}-\mu (\mu+s) t'-s\right)}{\mu  (\mu +s)^2} \; .
\end{eqnarray*}
Finally, after some calculations, we find
\begin{equation}
\label{navg}
\langle n (t') \rangle_s =\frac{k  \left(\mu+s e^{-(\mu+s)t' } \right) \left(\mu +s e^{-(\mu +s) (t-t')}\right)}{\mu  (\mu +s)^2}  \;\;\;\;\;\; 0 \le t' \le t \; .
\end{equation}
It is easy to see that 
$$
\langle n (t-t') \rangle_s = \langle n (t') \rangle_s
$$
which comes from the fact that $s$-ensemble is time-reversal symmetric. The result~(\ref{navg}) is also interesting from the 
point that in the long time limit the $s$-ensemble average of the number of proteins
at the initial and final times are given by
\begin{equation}
\langle n (0) \rangle_s = \langle n (t) \rangle_s = \frac{k}{\mu+s} \; .
\end{equation}
After a relaxation time $\tau=1/(\mu+s)$ the system falls into the TTI regime during which we have 
\begin{equation}
\langle n (t') \rangle_s = \frac{k \mu}{(\mu+s)^2}  \; .
\end{equation}

%%%%%%%%%%%%%%%%%
\begin{figure}[t]
\centering
\includegraphics[scale=0.55]{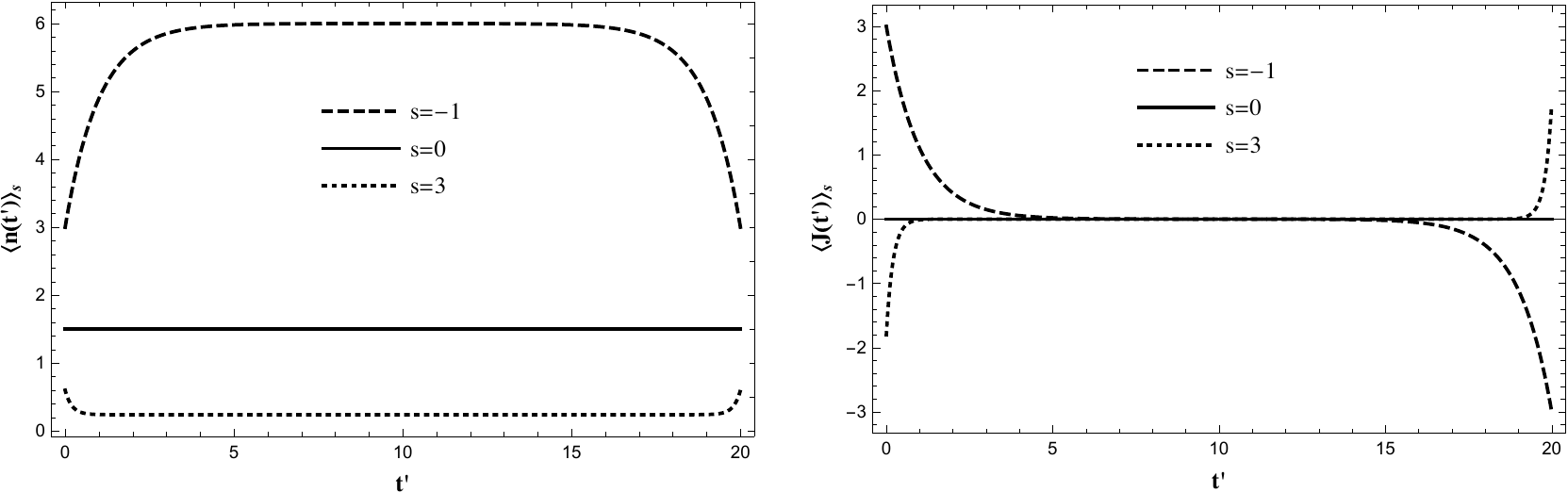}
\caption{\label{fig1} Plot of $\langle n(t') \rangle_s$ and $\langle J (t') \rangle_s$ as a function of $t'$ for $t=20$, $k=3$, $\mu=2$ and three different
values of $s$. At $s=0$ the system is in equilibrium hence both quantities are time independent. For $s \neq 0$ these quantities are constant only during 
the TTI regime.}
\end{figure}
%%%%%%%%%%%%%%%%%

As we mentioned, this model is in equilibrium in the long time limit i.e. the steady-state probability distribution $P(n)$ given by~(\ref{ss1}), 
besides the birth and death rates, satisfy the detailed balance condition. On the other hand, the average of the current of the proteins in the 
system (the current of birth minus the current of death) is zero because of the detailed balance. However, in the $s$-ensemble of
trajectories the average of this current is only zero during the TTI regime while it is non-zero during the initial and final transient regimes 
characterized by the relaxation time $\tau$. In order to clarify this, we have studied that average of the net current 
of proteins in the $s$-ensemble of number of proteins. The $s$-ensemble average of the net current is given by
\begin{eqnarray}
\frac{d}{dt}\langle n(t')\rangle_s&=&\langle J (t') \rangle_s = \langle J_{\mbox{\tiny Birth}} (t') - J_{\mbox{\tiny Death}} (t') \rangle_s \nonumber \\
&=& \frac{\langle 1 \vert e^{(t-t') H(s)} \hat{J} e^{t' H(s)} \vert \tilde{P}(s,0) \rangle}  {\langle 1 \vert \tilde{P}(s,t) \rangle } 
\end{eqnarray}
in which $J$ is a square matrix with the following elements 
\begin{eqnarray*}
\langle n+1 \vert \hat J \vert n \rangle = k\; , \\
\langle n-1 \vert \hat J \vert n \rangle = -n \mu \; .
\end{eqnarray*}
After some straightforward calculations one finds
\begin{equation}
\langle J (t') \rangle_s =  \frac{k s}{\mu +s} \left(e^{-(\mu+s) (t-t')}-e^{-(\mu+s)t'}\right)  \;\;\;\;\;\; 0 \le t' \le t \; .
\end{equation}
At three different times, including the TTI regime,  the asymptotic behavior of the current is given by 
\begin{eqnarray}
\langle J (0) \rangle_s &=&-\frac{k s}{\mu +s} \;\;\;\;\mbox{for}  \;\;\;\;  0= t' \ll t \; ,\\
\langle J (t') \rangle_s &=& 0 \;\;\;\;\mbox{for} \;\;\;\;  0 \ll t' \ll t \; ,\\
\langle J (t) \rangle_s &=& \frac{k s}{\mu +s}   \;\;\;\;\mbox{for}  \;\;\;\;  t'= t \to \infty  \; .
\end{eqnarray}
As can be seen the ensemble average of the net current is non-zero during the transient regimes while it is zero during the TTI regime as expected. 
In Fig.~\ref{fig1} we have plotted both $\langle n(t') \rangle_s$ and $\langle J (t') \rangle_s$ as a function of $t'$ for different values of $s$. As can be 
seen during the transient regimes both quantities are functions of time except $s=0$; however, there is a temporal regime where $\langle n(t') \rangle_s$ 
is a constant while $\langle J (t') \rangle_s$ is zero.  

We conclude the analysis of this model in the $s$-ensemble of trajectories by discussing the probability of observing a given configuration $n$ at time $t' \in [0,t]$
given that $s$ is fixed. This quantity is given by
\begin{equation}
P_s(n,t')=\frac{\langle 1 \vert e^{(t-t') H(s)} \vert n \rangle \langle n \vert  e^{t' H(s)} \vert \tilde{P}(s,0) \rangle}
                                           {\langle 1 \vert \tilde{P}(s,t) \rangle } \;.
\end{equation}
Straightforward calculations result in 
$$
P_s(n,t') =
\frac{e^{\frac{-k \left(s e^{-(\mu +s)t'}+\mu\right) 
   \left(s e^{-(\mu+s) (t-t' )}+\mu \right)}{\mu  (\mu +s)^2}}}{n!}
   \left(\frac{k \left(s e^{-(\mu +s)t'}+\mu\right) 
   \left(s e^{-(\mu+s) (t-t' )}+\mu \right)}{\mu  (\mu +s)^2}\right)^n \; .
$$
The fact that the $P_s(n,t') =P_s(n,t-t')$, which reveals the time-reversal symmetry of $s$-ensemble, can easily be seen. 
Now one can take the limit of this probability in the long time limit to find that 
\begin{eqnarray}
\label{Psnt}
\lim_{t \to \infty} P_s(n,t'=0) &=& \frac{e^{-\frac{k}{\mu+s}} }{n!} \left(\frac{k}{\mu +s}\right)^n  \nonumber \; ,\\
\lim_{t \to \infty} P_s(n,0 \ll t' \ll t) &=&  \frac{e^{-\frac{k \mu}{(\mu+s)^2}}}{n!} \left(\frac{k \mu}{(\mu +s)^2}\right)^n \; ,\\
\lim_{t \to \infty} P_s(n,t'=t) &=& \frac{e^{-\frac{k}{\mu+s}} }{n!} \left(\frac{k}{\mu +s}\right)^n \nonumber \; .
\end{eqnarray}
These can also be obtained from the properties of the left and the right eigenvectors of $H(s)$ associated with its largest eigenvalue, as
already explained in~(\ref{final}),~(\ref{initial}) and (\ref{tti}). More specifically, $\lim_{t \to \infty} P_s(n,0 \ll t' \ll t) $ can be obtained by
substituting the effective rates in the steady-state probability distribution function~(\ref{ss1}). 

\begin{figure}[t]
\centering
\includegraphics[scale=0.4]{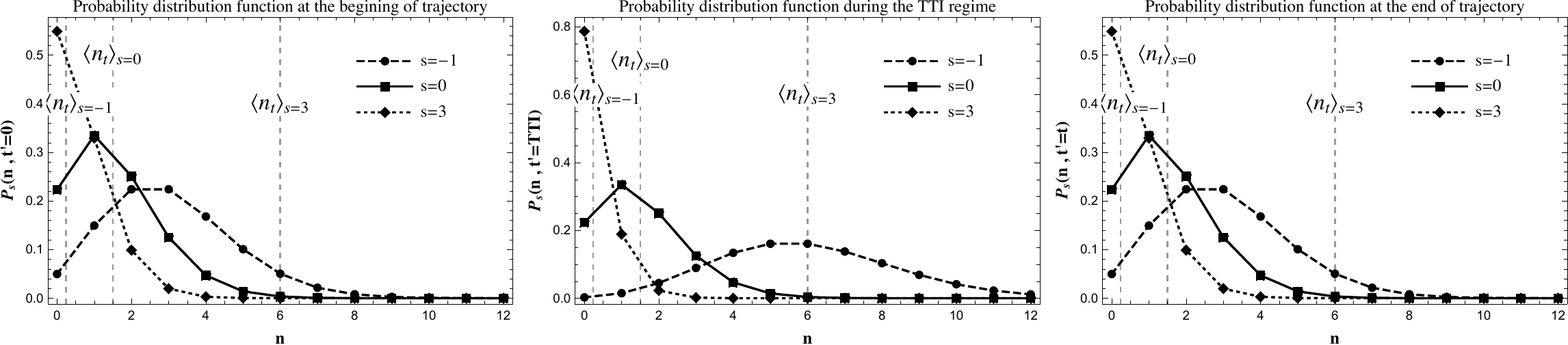}
\caption{\label{fig2} Plot of~(\ref{Psnt}) as a function of $n$ at $k=3$ and $\mu=2$. $\langle n_t \rangle_{s=-1}$, $\langle n_t \rangle_{s=3}$ and $\langle n_t 
\rangle_{s=0}$ correspond to $-\frac{d}{ds}\Lambda^{\ast}(s)$ at $s=-1$, $s=3$ and $s=0$ respectively. Note that a negative (positive) value of $s$ 
corresponds to a larger (smaller) than typical value of the observable.}
\end{figure}

In Fig.~\ref{fig2} we have plotted~(\ref{Psnt}) as a function of $t'$ for three different values of $s$. The diagrams
show which configuration is more probable in each case as $s$ is varied. As can be seen in~(\ref{Psnt}), the probability distributions at the 
beginning and the end of the trajectory are the same because of time-reversal symmetry. The average value of the number of proteins during 
the TTI regime is given by the first derivative of $\Lambda^{\ast}(s)$ with respect to $s$.

%%%%%%%%%%%%%%%%%
%%%%%%%%%%%%%%%%%

%%%%%%%%%%%%%%%%%%%%%%%%%%%%%%%%%%%%%%%%%%%%%%%%%%%%%%%%%%%%%%%%%%%%%%%%%
\subsection{Model 2}
%%%%%%%%%%%%%%%%%%%%%%%%%%%%%%%%%%%%%%%%%%%%%%%%%%%%%%%%%%%%%%%%%%%%%%%%%
Recent experiments have observed a feature of gene regulation which can be captured by defining stochastic processes in which protein
production often occurs in bursts resulting from many factors~\cite{ShR08,Fburst, Kburst}. In what follows we consider an effective model 
for protein production. We assume that protein production occurs with rates $k$, and burst sizes $n$ drawn from a state-dependent geometric 
distribution $b_n$ given by
\begin{equation}
\label{burstdist}
b_n=\frac{b^n}{(1+b)^{n+1}},\;\; n=0,1,2,\dots \;.
\end{equation}
The Master equation for the probability distribution $P(n,t)$ of having $n$ protein at time $t$ is
\begin{equation}
\frac{d}{dt}P(n,t)=k \sum_{r=0}^{n} b_r P(n-r,t) + \mu(n+1)P(n+1,t)-(k+n\mu)P(n,t) \; .
\end{equation}
The steady-state distribution is given by
\begin{equation}
P^{\ast}(n)=\frac{(\gamma)_n}{n!}  
  \Big(\frac{b}{1+b}\Big)^n (1+b)^{-\gamma}
\end{equation}
in which the symbol $(x)_n\equiv x(x+1)\cdots(x+n-1)$ is the ordinary Pochhammer symbol.  
The average number of proteins in the steady-state is also given by 
$$
\langle n \rangle = b \gamma \; .
$$
Let us now consider the number of proteins as a dynamical observable defines on the time-realizations of the process and investigate the 
fluctuations of this quantity. The modified generator for this process is given by
\begin{equation}
H(s)=
\left( \begin{array}{ccccccc}
-k+k b_0   & \mu          &     0                     &\cdots\\
k b_1    & -k+k b_0-\mu-s   &  2\mu                 &\cdots\\
k b_2    &     k b_1        &-k+k b_0-2\mu-2s      &\cdots\\
k b_3    &     k b_2          &       k b_1         &\cdots\\
\vdots   & \vdots             &    \vdots          &\ddots
\end{array} \right)\, \; .
\end{equation}
It turns out that the largest eigenvalue of $H(s)$ is given by
\begin{equation}
\Lambda^{\ast}(s)=\frac{-kbs}{\mu+s+bs} 
\end{equation}
and the right and left eigenvectors corresponding to this eigenvalue are also as follows
\begin{eqnarray}
 \vert \Lambda^{\ast} \rangle = \sum_{n=0}^{\infty} \frac{1}{n!}  \left(\frac{k}{\mu+s+bs}\right)_n
  \Big(\frac{b}{1+b}\Big)^n (1+b)^{\frac{-k}{\mu+s+bs}} \vert n \rangle\; , \\
\langle \tilde{\Lambda} ^{\ast} \vert= \sum_{n=0}^{\infty}  \Big(\frac{\mu}{\mu+s}\Big)^n   \langle n \vert \; .
\end{eqnarray}
Using the left eigenvector $\langle \tilde{\Lambda} ^{\ast} \vert$ the effective rates can be calculated using~(\ref{offelement})
\begin{eqnarray}
\langle n+r \vert  H_{\mbox{\tiny eff}}(s) \vert n \rangle &=& k b_r \Big(\frac{ \mu}{\mu +s}\Big)^r  \;,\\
\langle n-1 \vert  H_{\mbox{\tiny eff}}(s) \vert n \rangle &=& n (\mu +s) 
\end{eqnarray}
for $n,r=0,1,2,\dots$. Considering the number of proteins as a dynamical observable, its probability distribution in the long time limit
has a large deviation form which is given by
$$
P(n_t=\bar{n}) \asymp e^{-t I(\bar{n})}
$$
in which the rate function $I(\bar{n})$ can easily be calculated using~(\ref{ld})
\begin{equation}
I(n_t=\bar{n})=\frac{ kb+\bar{n} \mu-2\sqrt{ k b \bar{n}\mu}}{1+b} \; .
\end{equation}

%%%%%%%%%%%%%%%%%%%%%%%%%%%%%%%%%%%%%%%%%%%%%%%%%%%%%%%%%%%%%%%%%%%%%%%%%
\subsection{Model 3}
\begin{figure}[t]
\centering
\includegraphics[scale=1.5]{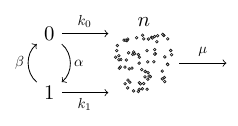}
\caption{\label{model3} A simple sketch of Model 3. }
\end{figure}

%%%%%%%%%%%%%%%%%%%%%%%%%%%%%%%%%%%%%%%%%%%%%%%%%%%%%%%%%%%%%%%%%%%%%%%%%
In the third example, we consider a stochastic process of gene expression from a promoter with $2$ internal states $i=0,1$. The promoter makes 
random transition from $0$ to $1$ with rate $\alpha$ and from $1$ to $0$ with rate $\beta$. In each state, a single mRNA is generated with rate 
$k_0$ ($k_1$) when the system is at the state $0$ ($1$)~\cite{Wbook}. This model is schematically drawn in Fig.~(\ref{model3})

%%%%%%%%%%%%%%%%%
%%%%%%%%%%%%%%%%%

In what follows we will show that it is possible to present a full description of the dynamical properties of the process. On the other hand,  we can 
discuss the fluctuations of the number of proteins in the long time limit. Considering the number of proteins as the dynamical observable the modified 
generator of the process in the complete orthonormal 
basis $\{ \vert i, n \rangle \} = \{ \vert 0,0\rangle, \vert 1,0 \rangle,\vert 0,1\rangle, \vert 1,1 \rangle,\dots \}$ is given by
\begin{equation}
H(s)=
\left( \begin{array}{ccccccc}
A_0   & D_1          &     0                 &   0                        &\cdots\\
B_1    & A_1   &  D_2           &    0                        &\cdots\\
0    &     B_1        &A_2     &   D_3           &\cdots\\
0    &    0          &       B_1     & A_3    &\cdots\\
\vdots            & \vdots                   &    \vdots                &\vdots          &\ddots
\end{array} \right)\,
\end{equation}
in which
$$
A_n=
\left(\begin{array}{cc}
-\alpha-k_0 -n (\mu+s)    &   \beta \\
\alpha & -\beta-k_1 -n (\mu+s)
\end{array}\right) ,\; D_n=
\left(\begin{array}{cc}
n \mu    &   0 \\
0           & n \mu 
\end{array}\right) 
$$
for $n=0,1,2,\dots$ and also
$$
B_1=
\left(\begin{array}{cc}
k_0     &   0 \\
0               & k_1  
\end{array}\right) \; .
$$
It turns out that the largest eigenvalue $\Lambda^\ast(s)$ and its corresponding left and right eigenvectors of $H(s)$ can be calculated exactly. Assuming 
the following form for the left eigenvector
\begin{equation}
\langle \tilde{\Lambda}^{\ast} \vert= \bigoplus_{n=0}^{\infty} (\frac{\mu}{\mu+s})^n\left(\begin{array}{cc} 
1 & \phi 
\end{array}\right)
\end{equation}
in which 
$$
\phi = \frac{\sqrt{\left(\left(k_0-k_1\right) s+(\alpha -\beta ) (\mu +s)\right)^2+4 \alpha  \beta  (\mu
   +s)^2}}{2 \alpha  (\mu +s)}  + \frac{\left(k_0-k_1\right) s+(\alpha -\beta ) (\mu +s)}{2 \alpha  (\mu +s)}
$$
one obtains the largest eigenvalue of the modified generator
\begin{eqnarray}
\fl
\Lambda^{\ast}(s)&&=
 \frac{\sqrt{\left(\mu  (\alpha +\beta )+
s \left(\alpha +\beta +k_0+k_1\right)\right)^2-4 s
   \left((\mu +s) \left(\alpha  k_1+\beta  k_0\right)+k_0 k_1 s\right)}}
   {2 (\mu+s)} \nonumber \\  \fl
   &&-\frac{\mu  ((\alpha +\beta ))+s \left(\alpha +\beta +k_0+k_1\right)}
   {2 (\mu+s)} \; .
\end{eqnarray}
As we mentioned earlier, the derivatives of $\Lambda^{\ast}(s)$ at $s=0$ generate the cumulants of the observable in the steady-state of the
process. It is easy to check the the average of the proteins in this model is given by 
$$\langle n_t \rangle_{s=0}=\frac{k_1 \alpha+k_0 \beta}{\mu (\alpha+\beta)}\; .$$

The right eigenvector corresponding to $\Lambda^{\ast}$ can also be calculated with more efforts. Let us consider
\begin{equation}
 \vert \Lambda ^{\ast} \rangle= \bigoplus_{n=0}^{\infty} \left(\begin{array}{c} 
\phi^{(0)}_n \\
  \phi^{(1)}_n 
\end{array}\right) \; .
 \end{equation}
Straightforward calculations show that the generating functions of $\phi^{(0)}_n $ and $\phi^{(1)}_n $ defined as
$$
f_0(x)  = \sum_{n=0}^{\infty} \phi^{(0)}_n  x^n\;, \;\;\; f_1(x)  = \sum_{n=0}^{\infty} \phi^{(1)}_n  x^n
$$
satisfy 
\begin{eqnarray*}
\Big( \mu-(\mu+s)x  \Big) f'_0(x)+\Big( k_0(x-1)-(\alpha+\Lambda^{\ast}(s)) \Big) f_0(x)+\beta f_1(x)&=&0 \; ,
\\
\Big( \mu-(\mu+s)x  \Big) f'_1(x)+\Big( k_1(x-1)-(\beta+\Lambda^{\ast}(s)) \Big) f_1(x)+\alpha f_0(x)&=&0 \; .
\end{eqnarray*}
We will show how to calculate $f_0(x)$, because $f_1(x)$ can easily be obtained by applying the following transformation to $f_0(x)$
$$
\alpha \rightleftarrows \beta\;\;\mbox{and}\;\; k_0 \rightleftarrows k_1\; .
$$
By defining a new variable $z$ as
$$
z\equiv -\mu+(\mu+s)x
$$
the equation governing $f_0(z)$ becomes
$$
-z^2 f''_0(z)+z(pz+q) f'_0(z) +(a z^2+b z+c) f_0(z)=0
$$
in which
\begin{eqnarray*}
\fl
p=\frac{k_0+k_1}{\mu +s} \; ,\\
\fl
a=-\frac{k_0 k_1}{(\mu +s)^2} \; ,\\
\fl
q= -\frac{\left(k_0+k_1\right) s+(\mu +s) (\mu +s+\alpha +\beta +2 \Lambda^\ast(s))}{\mu +s} \; ,\\
\fl
b= \frac{2 k_0 k_1 s+(\mu +s) \left(k_1 (\alpha +\Lambda^\ast(s) )+k_0 (\beta +\Lambda^\ast(s) +\mu +s)\right)}{(\mu +s)^2} \; ,\\
\fl
c= -\frac{k_0 k_1 s^2+s (\mu +s) \left(k_1 (\alpha +\Lambda^\ast(s))+k_0 (\beta +\Lambda^\ast(s) )\right)+\Lambda^\ast(s)  (\mu +s)^2 (\alpha +\beta
   +\Lambda^\ast(s) )}{(\mu +s)^2} \; .
\end{eqnarray*}
Now by choosing 
$$
f_0(z)=\mathcal{F}_0(z) e^{rz}z^t 
$$
in which
\begin{eqnarray*}
r&=&\frac{1}{2}(p-\sqrt{p^2+4a}) \;, \\
t&=&\frac{1}{2}(1+q+\sqrt{(1+q)^2+4c}) 
\end{eqnarray*}
we obtained the equation governing $\mathcal{F}_0(z)$ as follows
$$
\mathcal{A}(z) \mathcal{F}''_0(z)+\mathcal{B}(z) \mathcal{F}'_0(z) +\mathcal{C}(z) \mathcal{F}_0(z)=0
$$
in which
\begin{eqnarray*}
\mathcal{A}(z)&=& z \; ,\\
\mathcal{B}(z)&=& 1+\sqrt{(1+q)^2+4c}-z\sqrt{p^2+4a} \; ,\\
\mathcal{C}(z)&=& \frac{-1}{2}\Big(pq+2b+\sqrt{p^2+4a}(1+\sqrt{(1+q)^2+4c}\Big) \; .
\end{eqnarray*}
The final result is now given by a Hypergeometric function as follows 
$$
\mathcal{F}_0(z) =  \, _1F_1( \frac{pq+2b+\sqrt{p^2+4a}(1+\sqrt{(1+q)^2+4c})}{2\sqrt{p^2+4a}} ; 1+\sqrt{(1+q)^2+4c} ; z \sqrt{p^2+4a}) \; .
$$

Having the left eigenvector associated with the largest eigenvalue, we can calculate the effective rates as follows
\begin{eqnarray*}
\langle 0, n+1 \vert H_{\mbox{\tiny eff}}(s) \vert 0,n \rangle = \frac{ k_0 \mu}{\mu +s}  \;,\\
\langle 1, n+1 \vert H_{\mbox{\tiny eff}}(s) \vert 1,n \rangle = \frac{ k_1 \mu}{\mu +s}  \;,\\
\langle 0,n-1 \vert H_{\mbox{\tiny eff}}(s) \vert 0,n \rangle = n (\mu +s) \; ,\\
\langle 1,n-1 \vert H_{\mbox{\tiny eff}}(s) \vert 1,n \rangle = n (\mu +s)\; ,\\
\langle 0,n \vert H_{\mbox{\tiny eff}}(s) \vert 1,n \rangle = \beta \phi^{-1} \; ,\\
\langle 1,n \vert H_{\mbox{\tiny eff}}(s) \vert 0,n \rangle = \alpha \phi 
\end{eqnarray*}
for $n=0,1,2,3,\dots$. As can be seen, the effective rates are rescaled when compared to those in the original process. 

%%%%%%%%%%%%%%%%%%%%%%%%%%%%%%%%%%%%%%%%%%%%%%%%%%%%%%%%%%%%%%%%%%%%%%%%%
\subsection{Model 4}
%%%%%%%%%%%%%%%%%%%%%%%%%%%%%%%%%%%%%%%%%%%%%%%%%%%%%%%%%%%%%%%%%%%%%%%%%
The fourth example is similar to the third model except we have added burst. As in the previous model we consider a stochastic process of gene 
expression from a promoter with $2$ internal states $i=0,1$. The promoter makes random transition from $0$ to $1$ with rate $\alpha$ and from 
$1$ to $0$ with rate $\beta$. In each state, bursts of gene expression leading to the 
production of mRNAs occur with rate $k_0$ ($k_1$) when the system is at the state $0$ ($1$), and burst sizes $n$ drawn from a state-dependent 
distribution $b_n$ as introduced in~(\ref{burstdist}). This model can also be used to represent gene expression at the level of proteins. As in the 
previous models we are going to show that the largest eigenvalue of the modified generator besides the left eigenvector associated with this 
eigenvalue can be calculated exactly. These quantities are enough to provide us with a full description of the dynamical properties of the process. 

%%%%%%%%%%%%%%%%%
\begin{figure}[t]
\centering
\includegraphics[scale=.43]{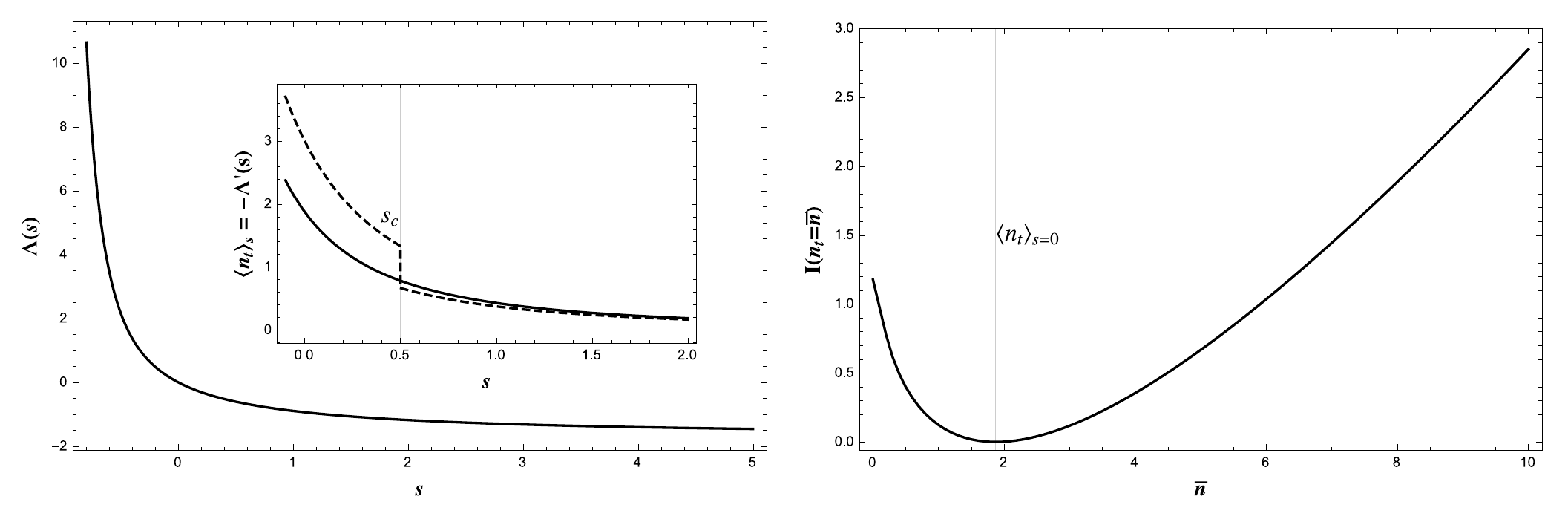}
\caption{\label{fig3} Plot of the largest eigenvalue $\Lambda^{\ast}(s)$ (left) and the large deviation function (or rate function) of
the model 4 (right) for $\alpha=0.5$, $\beta=1.5$, $k_0=3$, $k_1=6$, $\mu=2$ and $b=1$. The inset in the left figure shows the first
derivative of the largest eigenvalue for $\beta\neq 0$ (filled line) and $\beta=0$ (dashed line). }
\end{figure}
%%%%%%%%%%%%%%%%%

By considering the complete orthonormal basis $\{ \vert i, n \rangle \} = \{ \vert 0,0\rangle, \vert 1,0 \rangle,\vert 0,1\rangle, \vert 1,1 \rangle,\dots \}$ 
the modified generator of the process in the presence of burst is given by
\begin{equation}
H(s)=
\left( \begin{array}{ccccccc}
A_0   &D_1          &     0                 &   0                        &\cdots\\
B_1    & A_1   &  D_2           &    0                        &\cdots\\
B_2    &     B_1        &A_2     &  D_3           &\cdots\\
B_3    &   B_2          &   B_1     &A_3    &\cdots\\
\vdots            & \vdots                   &    \vdots                &\vdots          &\ddots
\end{array} \right)\,
\end{equation}
in which
$$
A_n=
\left(\begin{array}{cc}
-\alpha-k_0+k_0 b_0 -n (\mu+s)    &   \beta \\
\alpha & -\beta-k_1+k_1 b_0 -n (\mu+s)
\end{array}\right)  
$$
and also
$$
B_n=
\left(\begin{array}{cc}
k_0 b_n    &   0 \\
0               & k_1 b_n 
\end{array}\right) ,\;
D_n=
\left(\begin{array}{cc}
n \mu    &   0 \\
0           & n \mu 
\end{array}\right) 
$$
for $n=0,1,2,\dots$. It turns out that the largest eigenvalue and also the corresponding left eigenvector can be calculated as in the third model. 
Assuming the following form for the left eigenvector
\begin{equation}
\langle \tilde{\Lambda}^{\ast} \vert= \bigoplus_{n=0}^{\infty} (\frac{\mu}{\mu+s})^n \left(\begin{array}{cc} 
1 & \phi 
\end{array}\right)
\end{equation}
in which 
\begin{eqnarray*}
\phi &=& \frac{\sqrt{\left(b s \left(k_0-k_1\right) +(\alpha -\beta ) (b s+\mu +s)\right)^2
+4 \alpha  \beta  (b s+\mu +s)^2}}
   {2 \alpha  (b s+\mu +s)}  \\ \nonumber
   &+& \frac{b s \left(k_0-k_1\right) +(\alpha -\beta ) (b s+\mu +s)}
   {2 \alpha  (b s+\mu +s)}
\end{eqnarray*}
one obtains
\begin{eqnarray*}
\fl
\Lambda^{\ast}(s) &=&
 \frac{\sqrt{\left(b s \left(\alpha +\beta +k_0+k_1\right)+
 (\alpha +\beta ) (\mu +s)\right)^2-4 b s \left(b s k_0 k_1
 +\left(\alpha  k_1+\beta  k_0\right)(b s+\mu +s)\right)}}
   {2 (b s+\mu +s)} \nonumber \\
\fl
   &-&\frac{b s \left(\alpha +\beta +k_0+k_1\right)+(\alpha +\beta )
   (\mu +s)}
   {2 (b s+\mu +s)} \; .
\end{eqnarray*}
The first derivative of the largest eigenvalue at $s=0$ gives the average number of proteins in the steady-state
$$
\langle n_t \rangle_{s=0} =\frac{b(k_1 \alpha+k_0 \beta)}{\mu (\alpha+\beta)} \; .
$$
Note that for $k_0=k_1=k$, we recover the results of the second model i.e.
$$
\Lambda^{\ast}(s) = \frac{-kbs}{\mu+s+bs} \; .
$$
In Fig.~\ref{fig3} we have plotted the largest eigenvalue and the rate function calculated using the Legendre–Fenchel transformation~(\ref{ld}). 
Note that as $\beta$ approaches to zero the first derivative of the largest eigenvalue changes discontinuously at certain $s$. This indicates 
a first-order dynamical phase transition. 

In contrast to the previous cases, we have found that calculating the right eigenvector of $H(s)$ for an arbitrary $s$ is a formidable task. 
The effective rates of the effective dynamics can be calculated using~(\ref{offelement}) as for the previous models. The results are
\begin{eqnarray*}
\langle 0, n+r \vert H_{\mbox{\tiny eff}}(s) \vert 0,n \rangle =
k_0 b_r \Big(\frac{ \mu}{\mu +s}\Big)^r \; ,\\
\langle 1, n+r \vert H_{\mbox{\tiny eff}}(s) \vert 1,n \rangle =
k_1 b_r \Big(\frac{ \mu}{\mu +s}\Big)^r \; ,\\
\langle 0,n-1 \vert H_{\mbox{\tiny eff}}(s) \vert 0,n \rangle = n (\mu +s) \; ,\\
\langle 1,n-1 \vert H_{\mbox{\tiny eff}}(s) \vert 1,n \rangle = n (\mu +s) \; ,\\
\langle 0,n \vert H_{\mbox{\tiny eff}}(s) \vert 1,n \rangle = \beta \phi^{-1} \; , \\
\langle 1,n \vert H_{\mbox{\tiny eff}}(s) \vert 0,n \rangle = \alpha \phi 
\end{eqnarray*}
for $n,r=0,1,2,3,\dots$. As for the previous models, the effective rates of the effective dynamics are just rescaled in comparison to the
the reaction rates of the original process. 

%%%%%%%%%%%%%%%%%%%%%%%%%%%%%%%%%%%%%%%%%%%%%%%%%%%%%%%%%%%%%%%%%%%%%%%%%
\subsubsection{Generalization}
%%%%%%%%%%%%%%%%%%%%%%%%%%%%%%%%%%%%%%%%%%%%%%%%%%%%%%%%%%%%%%%%%%%%%%%%%
The previous model can be generalized to the one in which the promoter has more than two internal states. We assume that the promoter changes its 
internal state from $i$ to $j$ with the rate $\alpha_{i \to j}$ with $i,j=1,2,3,\dots,N$. The Master equation governing the 
time evolution of the probability distribution function for having $n$ proteins at time $t$ when the promoter is in the state $i$ is given by
\begin{eqnarray*}
\frac{d}{dt}P_i(n,t)&=&k_i \sum_{r=0}^{n} b_r P_i(n-r,t) + \mu(n+1) P_i (n+1,t) \\
                            &-&(k_i+\sum_{j \neq i} \alpha_{i \to j}+n\mu)P_i(n,t) \\                            
                            &+&\sum_{j \neq i} \alpha_{j \to i} P_j (n,t) \; .
\end{eqnarray*}
Considering the $s$-ensemble of number of proteins, the modified generator is given by
\begin{equation}
H(s)=
\left( \begin{array}{ccccccc}
A_0    & D_1          &     0    &   0                        &\cdots\\
B_1    & A_1          &D_2    &    0                        &\cdots\\
B_2    & B_1         &A_2     &D_3           &\cdots\\
B_3    &B_2          &B_1     &A_3    &\cdots\\
\vdots            & \vdots                   &    \vdots                &\vdots          &\ddots
\end{array} \right)\,
\end{equation}
in which the matrix elements of $N\times N$ matrices $A_n$, $B_n$ and $D_n$ are given by
\begin{eqnarray*}
(A_n)_{ij} =
\left\{\begin{array}{ll}
-\sum_{j' \neq i} \alpha_{i \to j'}-k_i+k_i b_0 -n (\mu+s) & i=j \\
\alpha_{j \to i} & i \neq j
\end{array}\right. \; ,\\
(B_n)_{ij} =
\left\{\begin{array}{cc}
k_i b_n & i=j  \\
0 & i \neq j
\end{array}\right. \; , \\
(D_n)_{ij} =
\left\{\begin{array}{cc}
n \mu & i=j 
\\
0 & i \neq j
\end{array}\right.
\end{eqnarray*}
for $i,j=1,2,3,\dots,N$. It turns out that, because of the special structure of $H(s)$, the largest eigenvalue and its corresponding left eigenvalue 
of $H(s)$, as an infinite-dimensional matrix, reduce to those of an $N \times N$ matrix. Considering the following left eigenvector  
\begin{equation}
\langle \tilde \Lambda^{\ast} \vert= \bigoplus_{n=0}^{\infty} \langle \tilde{\mathcal{X}}_n  \vert
\end{equation}
with
$$
\langle \tilde{\mathcal{X}}_n \vert=(\frac{\mu}{\mu+s})^n
\left(\begin{array}{cccc} 
\phi_0 &  \phi_1 & \dots & \phi_N
\end{array}\right)
$$
we only need to solve
\begin{equation}
\langle \tilde{\mathcal{X}}_0 \vert \tilde H(s)
= \Lambda^{\ast}(s) \langle \tilde{\mathcal{X}}_0 \vert
\end{equation}
in which $\tilde H(s)$ is defined as follows
\begin{eqnarray*}
(\tilde H(s))_{ij} = \left\{\begin{array}{cc}
-\sum_{j' \neq i} \alpha_{i \to j'}-k_i-n (\mu+s)+k_i(1+b-\frac{b \mu}{\mu+s})^{-1} & i=j  \\
\alpha_{j \to i} & i \neq j
\end{array}\right. \; .
\end{eqnarray*}
Now the effective rates can be calculated, after calculating $\phi_n$'s, as follows
\begin{eqnarray*}
\langle i, n+r \vert H_{\mbox{\tiny eff}}(s) \vert i,n \rangle =k_i b_r \Big(\frac{ \mu}{\mu +s}\Big)^r \; , \\
\langle i,n-1 \vert H_{\mbox{\tiny eff}}(s) \vert i,n \rangle = n (\mu +s) \; ,\\
\langle i,n \vert H_{\mbox{\tiny eff}}(s) \vert j,n \rangle = \alpha_{j \to i}   \frac{\phi_{i}}{\phi_j} 
\end{eqnarray*}
with $i,j=1,2,3,\dots,N$ and $n,r=1,2,3,\dots$.

Comparing with a recent work on stochastic gene expression conditioned on large deviations~\cite{HK17}, this generalization adds degradation to the process while
remaining integrable. As we saw, the conditioning-free effective process is represented by a process similar to the original process except 
that it has renormalized parameters~\cite{TJ15}.

%%%%%%%%%%%%%%%%%%%%%%%%%%%%%%%%%%%%%%%%%%%%%%%%%%%%%%%%%%%%%%%%%%%%%%%%%
\section{Concluding remarks}
%%%%%%%%%%%%%%%%%%%%%%%%%%%%%%%%%%%%%%%%%%%%%%%%%%%%%%%%%%%%%%%%%%%%%%%%%
In this paper, we studied four Markovian stochastic models of gene expression conditioned on large deviations of the population of proteins. Starting with 
a simple birth-death model, which is a continuous-time Markov process often used to study how the number of individuals in a population change through time, 
we showed that the model if fully integrable in the sense that the SCGF and also the corresponding left and right eigenvectors of the associated modified generator 
could be calculated exactly. The SCGF gives us, through the G\"artner–Ellis Theorem, the large deviation form of the probability distribution function of the number of 
proteins. Dynamical behavior of the system also obtained using the probabilistic concepts of the left and right eigenvectors associated with the SCGF.

We gradually generalized the simple birth-death process to the stochastic processes which were more realistic in the realm of gene expression. We showed
that they were still integrable and fluctuations of the dynamical observable could be traced using the large deviation theory. Finally, we showed that the modeling
of gene expression as a Batch Markovian Arrival Process (BMAP) studied in~\cite{HK17} could be generalized to a model which includes degradation and that this generalization
does not alter the integrability of the system. 

As it has been explained in~\cite{ZRPsym}, the SCGF is not always given by the largest eigenvalue of the modified generator. This might happen in the systems with 
infinite-dimensional configuration space. We have checked that, despite the configuration space is infinite-dimensional, the SCGF is still given by the 
largest eigenvalue of the modified generator.   

%%%%%%%%%%%%%%%%%%%%%%%%%%%%%%%%%%%%%%%%%%%%%%%%%%%%%%%%%%%%%%%%%%%%%%%%%
\section*{Acknowledgment }
We would like to thank Prof. Rahul Kulkarni for his suggestions through an email correspondence.  
%%%%%%%%%%%%%%%%%%%%%%%%%%%%%%%%%%%%%%%%%%%%%%%%%%%%%%%%%%%%%%%%%%%%%%%%%

%%%%%%%%%%%%%%%%%%%%%%%%%%%%%%%%%%%%%%%%%%%%%%%%%%%%%%%%%%%%%%%%%%%%%%%%%
\section*{References}
%%%%%%%%%%%%%%%%%%%%%%%%%%%%%%%%%%%%%%%%%%%%%%%%%%%%%%%%%%%%%%%%%%%%%%%%%

\end{document}